\documentstyle[12pt,epsfig]{article}

\newlength{\dinwidth}
\newlength{\dinmargin}
\begin{document}
\def\bold#1{\setbox0=\hbox{$#1$}%
     \kern-.025em\copy0\kern-\wd0
     \kern.05em\copy0\kern-\wd0
     \kern-.025em\raise.0433em\box0 }
\def\slash#1{\setbox0=\hbox{$#1$}#1\hskip-\wd0\dimen0=5pt\advance
       \dimen0 by-\ht0\advance\dimen0 by\dp0\lower0.5\dimen0\hbox
         to\wd0{\hss\sl/\/\hss}}
\newcommand{\be}{\begin{equation}}
\newcommand{\ee}{\end{equation}}
\newcommand{\bea}{\begin{eqnarray}}
\newcommand{\eea}{\end{eqnarray}}
\newcommand{\nn}{\nonumber}
\newcommand{\dd}{\displaystyle}
\newcommand{\bra}[1]{\left\langle #1 \right|}
\newcommand{\ket}[1]{\left| #1 \right\rangle}
\newcommand{\spur}[1]{\not\! #1 \,}
\thispagestyle{empty}
\rightline{ DCPT/01/34 - IPPP/01/17}
\rightline{ BARI-TH/413-2001}
\rightline{April 2001}
\vspace*{2cm}
\begin{center}
  \begin{LARGE}
  \begin{bf}
Probing the structure of $f_0(980)$ through radiative $\phi$ decays
  \end{bf}
  \end{LARGE}
\end{center}
\vspace{8mm}
\begin{center}
  \begin{large}
 F. De Fazio$^{a}$ and M.R. Pennington$^{b}$
  \end{large}
\end{center}
  \vspace{1cm}

\begin{center}
\begin{it}
$^{a}$ Istituto Nazionale di Fisica Nucleare, Sezione di Bari, Italy \\
$^{b}$ Institute for Particle Physics Phenomenology, University of
Durham,\\
DH1 3LE, Durham, UK
\end{it}
\end{center}
\begin{quotation}
\vspace*{1.5cm}
\begin{center}
{\bf
  Abstract}\\
\end{center}
We consider the radiative transition $\phi \to f_0 \gamma$, which is a
sensitive probe of the nature of the $f_0(980)$ particle. Using the QCD
sum-rule technique, we estimate the branching ratio of such decay mode to
be:
${\cal B}(\phi \to f_0 \gamma)=(2.7 \pm 1.1) ~ 10^{-4}$, in
fair agreement with present experimental data. As for the structure
of the $f_0$, the result suggests a sizeable $s {\bar s}$ component;
however, this result  does not exclude the possibility of
further components and allows a more complex structure than indicated
by the naive quark model.
\noindent
\vspace*{0.5cm}
\end{quotation}
\newpage
\baselineskip=6.5mm
\vspace{2cm}
\noindent
The quark model provides a rather good description of hadrons,
which
fit  into suitable multiplets reasonably well.  
In its simplest version the model then
interprets  mesons as pure $q{\overline q}$ states. Scalar mesons present a
remarkable
exception to this successful scheme. Indeed, the nature of these mesons is not
established yet \cite{rassegne}. There are more scalars than can fit
into one quark model multiplet. Consequently, some of these states
could be either glueballs or admixtures of quark and gluonic states, or belong to multiquark multiplets. A
particular feature of some of these particles is that they appear to be  rather wide 
\cite{isgur,au,janssen,oller}.
They have very short lifetimes
and large couplings to hadronic channels, such as $K {\overline K}$ or $\pi \pi$.
This might  suggest  that they can be identified as composite
systems of hadrons, or
that they spend an appreciable part of their lifetimes as such states.
This could be the result of  hadronic dressing, whereby the
strong interaction enriches a $q{\overline q}$ state with other components such
as $\ket{K {\overline K}}$, $\ket{\pi \eta}$, etc.
 Such a  viewpoint could also
explain why the scalar mesons seem to contradict the OZI rule. Since the two
mesons composing the state in which they spend much of their
lifetime may readily annihilate to $q{\overline q}$, leading to a subsequent OZI allowed
decay.

In this letter, we focus on
the structure of the $f_0(980)$ and the possibility of gleaning information
about this from radiative $\phi$ decays.
According to the quark model, the $f_0(980)$ should be an $s{\overline s}$
state, an interpretation supported in Refs. \cite{tornqvist,
roos,scadron}. However, this does not explain its mass degeneracy
with the $a_0(980)$, that should be a $({u {\overline u} -d {\overline
d})/\sqrt{2}}$ state.
There are also suggestions that the $f_0(980)$ could be a four quark
$qq{\overline{qq}}$ state \cite{jaffe}.
In this case, it could either be
nucleon-like \cite{ivan}, {\it i.e.} a bound state of quarks  with symbolic
quark structure ${s{\overline s}({ u {\overline u}+d {\overline d})/
\sqrt{2}}}$ 
\footnote{Within the same
framework the isovector partner $a_0(980)$ is written as
 $s {\overline s}( u {\overline u}+d {\overline d}) / \sqrt{2}$.}, 
or deuteron-like, {\it i.e.} a bound
state of hadrons, which is usually referred to as a $K {\overline K}$ molecule
\cite{isgur,closebook,kaminski,shev}. In the former of these two
possibilities, the mesons
are treated as point-like objects, while in the latter they should be
considered as extended objects. 
Some objections have been raised against the $K{\overline K}$ 
molecular model \cite{isgur,morgan}. In particular, such an
interpretation requires a width smaller than the binding energy of
the molecule itself, which has been estimated  to be $\epsilon
\simeq 10-20$ MeV \cite{isgur}, in contrast to the measured width lying
in the range $40-100$ MeV \cite{pdg}.

Various ways have been suggested of clarifying the
situation,
such as the analysis of the $f_0 \to \gamma \gamma$
decay \cite{twophoton,boglione}
 or of the ratio $\displaystyle{\Gamma(\phi \to a_0 \gamma) \over
\Gamma(\phi \to f_0 \gamma)}$ \cite{closebook}. In the
naive quark model,  for example, it is expected that 
${\cal B}(\phi \to f_0 \gamma)$ and 
${\cal B}(\phi \to a_0 \gamma)$ would differ by a factor of 10.
Moreover the rate for $\phi \to f_0 \gamma$
may distinguish among the different possibilities \cite{closebook},
since,
according to the existing
theoretical estimates, the expected branching ratio would be as high as
$10^{-4}$ in the $qq{\overline {qq}}$ case, ${\cal O}(10^{-5})$ in the
$s{\overline s}$ case.
 For a $K{\overline K}$ molecule, the branching ratio clearly depends
on its size. For a compact state this is $\sim 7 \cdot 10^{-5}$, while
for a diffuse, deuteron-like system, it is down below $10^{-5}$
\cite{closebook}.

From the experimental point of view, the PDG value \cite{pdg}:
\be
{\cal B}(\phi \to f_0 \gamma)\,=\,(3.4 \pm 0.4)~10^{-4}
\ee
stems from averaging the results of the CMD2 \cite{akh}
and SND \cite{snd} collaborations, analysing $\pi^+ \pi^- \gamma$, $\pi^0
\pi^0 \gamma$ and $5 \gamma$ final states. What is more, a significant
improvement is expected at the $\phi$ factory
DA$\Phi$NE \cite{book}, where the first results give: 
\be
\nonumber
{\cal B}(\phi \to
f_0 \gamma\to \pi^0 \pi^0 \gamma)=(0.81 \pm 0.09 ({\rm stat}) \pm 0.06({\rm syst}))\times 10^{-4} \nonumber
\ee
and 
\be
\nonumber
{\cal B}(\phi \to
f_0 \gamma\to \pi^- \pi^+ \gamma)< 1.64 \times 10^{-4}
\nonumber 
\ee
 at $90\%$ C.L. \cite{kloe}.

The present letter is devoted to analysis of the radiative decay $\phi
\to f_0 \gamma$ using
 QCD sum-rules
\cite{shifman}, which we previously applied  to the radiative $\phi$
transitions to $\eta,\eta^\prime$ \cite{noi}. 
That the $f_0(980)$ couples significantly through $s{\overline s}$
components has long been known \footnote{And noticed more recently by 
Delbourgo et al. \cite{liu} for $\phi \to f_0 \gamma$.} 
from its appearance as a peak
in $J/\psi\to\phi f_0$ \cite{gidal} and $D_s\to\pi f_0$ \cite{anjos}, as
discussed in Refs. \cite{au}, and in more detail in \cite{morgan1}.
Our calculation relies on the assumed coupling of the $f_0$ to
the scalar $s{\overline s}$ density. As a preliminary, we evaluate
the strength of this coupling using two point QCD sum-rules.
The result will then  be exploited in the three point QCD sum-rule
evaluation of the relevant quantity needed to compute ${\cal
B}(\phi \to f_0 \gamma)$.

The coupling of the $f_0(980)$ to the scalar current $J^s={\overline s}s$
can be
parametrized in terms of a constant ${\tilde f}$:
\be
\bra{0}  J^s \ket{f_0(p)}= m_{f_0} ~{\tilde f} \;. \label{ftilde}
\ee
\noindent In order to compute this parameter by QCD sum-rules, we consider
the two-point correlator:
\be
T(q^2)=i ~\int d^4 x e^{iq \cdot x} \bra{0} T[ J^s(x) J^{s
\dagger}(0)]\ket{0} \;,
\label{cor}
\ee
\noindent which is given by the dispersive representation:
\be
T(q^2)={1 \over \pi} \int_{4 m_s^2}^\infty \, ds\;{\rho(s) \over s-q^2}
+{\rm subtractions} \;.
\label{disp}
\ee
\noindent In the region of low values of $s$, the physical spectral
density
contains a $\delta-$function term corresponding, in the small width
approximation, to the coupling of
the $f_0$ to the scalar current. Picking up this contribution and  
 dropping possible subtractions which we discuss later, we can write:
\be
T(q^2)={m_{f_0}^2~{\tilde f}^2 \over m_{f_0}^2-q^2}+ {1 \over \pi}
\int_{s_0}^\infty\,ds\;
{\rho^{had}(s) \over s-q^2} \;,
\ee
\noindent  assuming that the contribution of higher
resonances and continuum of states start from an effective threshold
$s_0$.
On the other hand, the correlator $T(q^2)$  can be computed in QCD for
large Euclidean values of $q^2$, by using the Operator Product Expansion (OPE)
to expand the $T$-product in Eq.~(\ref{cor})
as the sum of a perturbative contribution plus
non-perturbative terms which are proportional to vacuum expectation values
of
quark and gluon gauge-invariant operators of increasing dimension, the so
called {\it vacuum condensates}. In practice, only a few condensates are
included, the most important contributions coming from  the dimension 3
$<{\overline q} q>$ and dimension 5 $<{\overline q}g \sigma  G q>$. 

In the QCD expression for the two-point correlator considered, the
perturbative term can also be written dispersively, so that:
\be
T^{QCD}(q^2)={1 \over \pi} \int_{4 m_s^2}^\infty\,ds\; {\rho^{pert}(s)
\over
s-q^2} + d_3 <{\overline s } s> +d_5<{\overline s}g \sigma  G s>+...
\label{qcd} \; ,
\ee
\noindent where the spectral function $\rho^{pert}$ and the coefficients
$d_3$, $d_5$ can be computed in QCD.
The next step consists in assuming quark-hadron duality, which amounts to
the claim that the physical and the perturbative spectral densities
give the same result
when integrated appropriately above some $s_0$. This leads to the
sum-rule:
\be
{m_{f_0}^2~{\tilde f}^2 \over m_{f_0}^2-q^2}
={1 \over \pi} \int_{4 m_s^2}^{s_0}\,ds\;
{\rho^{pert}(s) \over s-q^2} + d_3 <{\overline s } s> +d_5<{\overline s}g
\sigma Gs>+ \dots \label{sr}
\ee
\noindent This expression 
can be improved by applying
to both sides of Eq.~(\ref{sr}) a  Borel transform, defined as follows:
\be
{\cal B} [{\cal F}(Q^2)]=lim_{Q^2 \to \infty, \; n \to \infty, \; {Q^2 \over
n}=M^2}\;
{1 \over (n-1)!} (-Q^2)^n \left({d \over dQ^2} \right)^n {\cal F}(Q^2) \; ,
\label{tborel}
\ee
\noindent where ${\cal F}$ is a generic function of $Q^2=-q^2$. The application
of
such a procedure to the sum-rules amounts to exploiting the following
result:
\be
{\cal B} \left[ { 1 \over (s+Q^2)^n } \right]={\exp(-s/M^2) \over 
(M^2)^n\ (n-1)!} \; , \label{bor}
\ee
\noindent where $M^2$ is known as the Borel parameter. This operation
improves the
convergence of the series in the OPE by factorials in $n$ and, for suitably
chosen
values of $M^2$, enhances the contribution of low lying states. Moreover,
since the Borel transform of a polynomial vanishes, it is correct to
neglect subtraction terms in Eq.~(\ref{disp}), which are polynomials
in $q^2$.
The final sum-rule reads:
\bea
&& m_{f_0}^2~{\tilde f}^2 \exp\Bigg(-{m_{f_0}^2 \over M^2}\Bigg) = 
{3 \over 8 \pi^2}
\int_{4 m_s^2}^{s_0} ds~s ~\Bigg( 1-{4 m_s^2 \over s}\Bigg)^{3/2}\
\exp\Bigg(-{s \over M^2}\Bigg)
\nonumber \\
&&+ m_s\ \exp\Bigg(-{m_s^2 \over M^2}\Bigg) \Bigg[ <{\overline s} s> 
\Bigg(3 + {m_s^2\over M^2} +{m_s^4 \over M^4} \Bigg)+ 
<{\overline s}g \sigma G s> { 1
\over M^2} \Bigg( 1- {m_s^2 \over 2 M^2} \Bigg) \Bigg]
\label{finsr} \; .
\eea
\noindent In the  numerical evaluation of Eq.~(\ref{finsr})
we use $<{\overline s}s>=0.8 <{\overline q}q>$,
$<{\overline q}q>=(-0.24)^3$ GeV$^3$, $<{\overline s}g \sigma G
s>=0.8~$GeV$^2 <{\overline s}s>$,
$m_{f_0}=0.980$ GeV.
The strange quark mass is chosen in the
range $m_s=0.125-0.160$ GeV,
obtained in the same QCD sum-rule framework \cite{noistrano}.
The threshold is chosen
below a possible  $f_0(1370)$ pole and varied between
$s_0=1.6-1.7$ GeV$^2$.
Since the Borel parameter has no physical meaning, we look for a range of
its values (``stability window'')
where the sum-rule is almost independent on $M^2$. Such a window is
usually sought
in a restricted interval of values of the Borel parameter chosen
 by requiring that the perturbative
contribution is at least 20$\%$ of the continuum and
additionally
requiring that the perturbative term is greater than the
non-perturbative contribution.
\begin{figure}                                                      
\begin{center}                                                      
\vspace{-1.5cm}
\epsfig{file=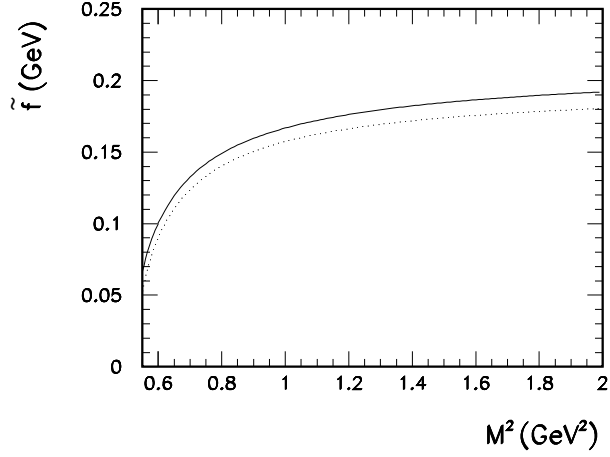,height=10cm} 
\vspace{-1.5cm}                                                     
\caption{Coupling of the $f_0$ to the scalar current as a function  
of the Borel parameter $M$, for $m_s=0.140$ GeV. The solid  curve   
corresponds to the higher threshold $s_0=1.7\;{\rm GeV}^2$, the     
dashed curve corresponds to $s_0=1.6\;{\rm GeV}^2$.\label{figft}} 
\end{center}                                                        
\end{figure}
The stability window for $M^2$ is selected in
$[1.2,2]$ GeV$^2$, as seen in Fig. \ref{figft}, where,
taking into
account  the uncertainty on $m_s$, we obtain the coupling:
\be
{\tilde f}=(0.180 \pm 0.015)\; {\rm GeV} \label{ftris} \; .
\ee
\noindent This result will be used  in the analysis of the decay $\phi
\to f_0 \gamma$ as we shall see in the following.

The relevant matrix element describing the transition $\phi
\to f_0 $ induced by a strange vector current $J_\mu={\overline s} \gamma_\mu s$, 
can be parameterized as follows:
\bea
\bra{ f_0(q_2)} J_\mu \ket{\phi (q_1, \epsilon_1)} &=& F_1(q^2)~
(q_1
\cdot q_2) ~
\epsilon_{1 \mu} + F_2(q^2)~ (\epsilon_1 \cdot q_2)~ (q_1+q_2)_\mu
\nonumber \\
&+& F_3(q^2) ~(\epsilon_1 \cdot q_2)~ q_\mu
\label{matrix}
\eea
\noindent where $q=q_1-q_2$. 
In order to consider the radiative decay $\phi \to f_0 \gamma$, one needs
the amplitude
\be
{\cal A}(\phi (q_1, \epsilon_1)\to f_0(q_2)\gamma(q, \epsilon))=-{1 \over
3}~ e~\epsilon^{* \mu}~ \big[ F_1(0)~ (q_1
\cdot q_2)~
\epsilon_{1 \mu} + F_2(0)~ (\epsilon_1 \cdot q_2)~ (q_1+q_2)_\mu
\big] \;, \label{ampl}
\ee
\noindent where the charge of the strange quark has been explicitly
written.
Eq.~(\ref{ampl}) shows that only two of the three form factors
appearing in Eq.~(\ref{matrix}) are actually needed.
Furthermore, gauge invariance requires that $q^\mu \cdot \big[ F_1(0)~
(q_1 \cdot q_2)
\epsilon_{1 \mu} + F_2(0)~ (\epsilon_1 \cdot q_2)~ (q_1+q_2)_\mu
\big]=0$, which
 relates the values of $F_1$ and $F_2$ at $q^2=0$:
\be
F_2(0)=F_1(0)\,{ m_\phi^2+m_{f_0}^2 \over 2 ( m_\phi^2-m_{f_0}^2 )} \;.
\label{gaugein}
\ee
\noindent In terms of $F_1(0)$, the rate for the process we consider becomes:
\be
\Gamma(\phi \to f_0 \gamma)=\alpha~ [F_1(0)]^2~ {( m_\phi^2-m_{f_0}^2 )
(m_\phi^2+m_{f_0}^2)^2 \over 216~ m_\phi^3} \;.
\label{rate}
\ee
\noindent 
Three-point QCD sum-rules can be applied to evaluate the form
factor $F_1(q^2)$.
We  consider the three-point function:
\be
\Pi_{\mu \nu}(q_1^2,q_2^2,q^2)=i^2 \int d^4 x~ d^4 y~ e^{-i q_1 \cdot x}\,
e^{i q_2 \cdot y} \bra{0} T[J^s(y) J_\nu(0) J_\mu(x)]\ket{0}
\label{cor-phi}
\ee
where $J^s$ has been defined above and $J_\nu={\overline s} \gamma_\nu
s$ is
the vector current. The correlator Eq.~(\ref{cor-phi}) can be
written in terms of invariant structures as follows:
\be
\Pi_{\mu \nu}(q_1^2,q_2^2,q^2)=\Pi(q_1^2,q_2^2,q^2)\,g_{\mu \nu}+
\Pi_1(q_1^2,q_2^2,q^2)\,q_{1 \mu} q_{1\nu} + \cdots
\label{pi}
\ee
\noindent and a QCD sum-rule can be built up for the structure
$\Pi(q_1^2,q_2^2,q^2)$.
The method closely follows the one described for the two-point sum-rule.
We assume $\Pi(q_1^2,q_2^2,q^2)$ obeys a dispersion relation
in both
the variables $q_1^2,q_2^2$:
\be
 \Pi(q_1^2,q_2^2,q^2)={1 \over \pi^2} \int ds_1 \int ds_2\,
{\rho(s_1,s_2,q^2) \over (s_1 -q_1^2)(s_2-q_2^2)} \; ,
\ee
\noindent with possible subtractions.
Such a representation is true at each order in perturbation theory and, as
is
standard in QCD sum rule analyses, it is assumed to hold in general.
In this case 
the spectral function contains,
for low values of $s_1$,
$s_2$, a double $\delta-$function corresponding to the transition $\phi
\to
f_0$. Extracting this contribution, we can write:
\be
 \Pi(q_1^2,q_2^2,q^2)=-{m_{f_0}\ {\tilde f}\ m_\phi\ f_\phi\ F_1(q^2)(q_1
\cdot q_2)\over
(m_\phi^2-q_1^2)(m_{f_0}^2-q_2^2)} +{1 \over
\pi^2}\int_{s_0^\prime}^\infty
ds_1 \int_{s_0}^\infty d s_2\, {\rho^{had}(s_1,s_2,q^2) \over (s_1
-q_1^2)(s_2-q_2^2)} \; ,
\ee
\noindent where subtractions are neglected as later they will vanish on
taking a Borel transform.
The parameter ${\tilde f}$ appearing in the previous equation is just the
coupling  of the
$f_0$ to the scalar current, computed  previously.
Deriving an OPE-based QCD expansion for $\Pi$ for large and negative
$q_1^2$, $q_2^2$ and $q^2$,  one can write:
\be
\Pi(q_1^2,q_2^2,q^2)={1 \over \pi^2}\int_{4 m_s^2}^\infty d s_1 \int_{4
m_s^2}^\infty d s_2
{\rho^{pert}(s_1,s_2,q^2) \over (s_1 -q_1^2)(s_2-q_2^2)}
+ c_3 <{\overline s}
s > +c_5 < {\overline s} g \sigma G s
> +...\;\;.
\ee
Invoking quark-hadron global duality as before, we arrive at the sum-rule:
\bea
{m_{f_0}\ {\tilde f}\ m_\phi\ f_\phi\ F_1(q^2)\ (q^2-m_\phi^2-m_{f_0}^2) \over 2
(m_\phi^2-q_1^2)(m_\eta^2-q_2^2)} &=&
{1 \over \pi^2}\int_D d s_1  d s_2\,
{\rho^{pert}(s_1,s_2,q^2) \over (s_1 -q_1^2)(s_2-q_2^2)} \nonumber \\
&+& c_3 <{\overline s }
s>+c_5 <{\overline s} g \sigma G s>+...\;\;, \label{4.7}
\eea
where the domain $D$ should now also satisfy the kinematical
constraints specified below.
After a double Borel transform in the variables $-q_1^2$
and $-q_2^2$, we obtain:
\bea
&&{1\over 2}\ m_{f_0} {\tilde f}  m_\phi  f_\phi (q^2-m_\phi^2-m_{f_0}^2)\,
F_1(q^2)\
\exp\Big(-{m_{\phi}^2 \over M_1^2} - {m_{f_0}^2 \over M_2^2}\Big)\, = 
\nonumber \\
&&{\hspace{1.5cm}}{1 \over \pi^2} \int_D ds_1 ds_2 \exp\Big(-{s_1 \over M_1^2}
-{s_2 \over M_2^2}\Big)\ \rho^{pert}(s_1,s_2)
 \label{srb} \\
&&{\hspace{1.5cm}}+ \exp\Big(-{m_s^2\over M_1^2} - {m_s^2 \over M_2^2}\Big)\
\Bigg\{ <{\overline s} s> \Bigg[\ q^2 +2 m_s^2-{m_s^2 q^2 \over M_1^2}
\nonumber\\&& {\hspace{6.5cm}}+
{m_s^2 q^2 (2 m_s^2-q^2) \over 2 M_1^2 M_2^2} +{m_s^4 q^2 \over 2} \left(
{1 \over M_1^4}+{1 \over M_2^4} \right)\Bigg] \nonumber \\
&&{\hspace{1.5cm}}+ <{\overline s } g \sigma G s> \Bigg[ -{1 \over 3}+
{q^2 -m_s^2 \over 3 M_1^2}+{2 q^2+m_s^2 \over 3 M_2^2}\nonumber \\
&& {\hspace{6.5cm}} -{q^2 (5 m_s^2-2 q^2) \over 6 M_1^2
M_2^2}-{m_s^2 q^2 \over 4 M_2^4} -{m_s^2 (3 q^2 +m_s^2) \over 12 M_1^4}
\Bigg] \Bigg\} \;,\nonumber
\eea
\noindent where:
\bea
\rho^{pert}(s_1,s_2)=\, {3 m_s\over 4} \Big\{ ( 4 m_s^2+ s_1-s_2+q^2 )
\Big[ \Big(s_1+s_2-q^2\Big)^2- 4s_1 s_2 \Big] + 4 q^2 s_1 s_2
\Big\}/&&
\nonumber \\
&&{\hspace{-6.cm}} \Big[ (s_1+s_2-q^2)^2
- 4s_1 s_2 \Big]^{3/2} \;. \label{rhopert}
\eea
\noindent
The integration domain $D$ over the variables $s_1,s_2$ depends on
the value of $q^2$.
For $(-q^2)> s_0-4 m_s^2$, $D$ is specified by:
  $(s_2)_- \le s_2 \le s_0$   $4
m_s^2 \le s_1 \le s_0^\prime$; while,
for $(-q^2)< s_0-4 m_s^2$
$D$ is bounded by:
$(s_2)_- \le s_2 \le (s_2)_+ $ if
$4 m_s^2 \le s_1 \le
(s_1)_-$ and
$(s_2)_- \le s_2 \le s_0$ if $(s_1)_- \le s_1 \le
s_0^\prime$, with:
$(s_2)_\pm=\left[2 m_s^2 q^2+(2 m_s^2-q^2) s_1 \pm \sqrt{s_1 q^2 (q^2-4
m_s^2)(s_1-4 m_s^2)} \right] / 2 m_s^2$
and $(s_1)_\pm=\left[ 2 m_s^2 q^2+(2 m_s^2-q^2) s_0 \pm \sqrt{s_0 q^2
(q^2-4 m_s^2)(s_0-4 m_s^2)} \right]/ 2 m_s^2$.

\begin{figure}
\begin{center}
\vspace{-1.5cm}
\epsfig{file=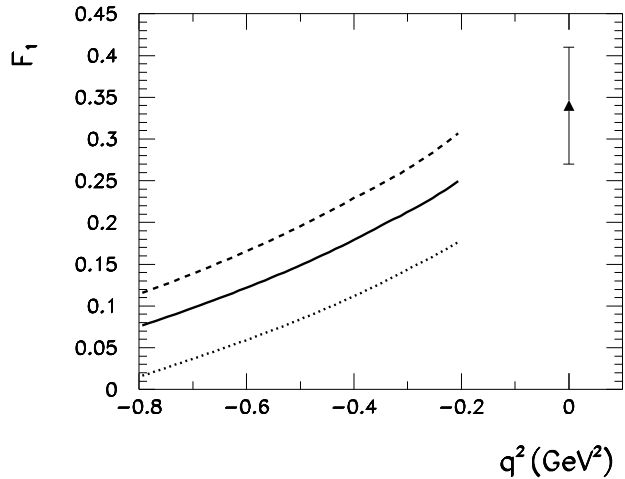,height=10cm} 
\vspace{-1.5cm}
\caption{Form factor $F_1(q^2)$. The dashed and dotted lines are
the highest and the lowest curves obtained 
varying the set of parameters entering in the sum rule (\ref{srb}). The
isolated point on the right is the result of an extrapolation. The result 
(\ref{f1}) corresponds to the central point on the right obtained by
extrapolating the solid curve.\label{figf1}} \end{center}
\end{figure}

Since we  consider the form-factor $F_1(q^2)$ for arbitrary negative
values of
$q^2$, we could perform a double Borel transform in the two
variables $Q_1^2=-q_1^2$ and $Q_2^2=-q_2^2$, which allows us to remove
single poles in the $s_1$ and $s_2$ channels
from the  sum-rule. Our procedure
 is therefore to compute the form-factor $F_1(q^2)$ and
then to extrapolate the result to $q^2=0$.
In the numerical analysis we use:
$ m_\phi=1.02$ GeV,
$f_\phi=0.234$ GeV (obtained from the experimental datum on the decay to
$e^+ e^-$ \cite{pdg}). We compute the result for  two values of
the $\phi$ threshold: $s_0^\prime=1.8,1.9$ GeV$^2$.
$s_0$ coincides with the $f_0$ threshold chosen as  for the
two point function.
 The extrapolation to $q^2=0$  shown in Fig.~\ref{figf1} gives:
\be
F_1(0)=0.34 \pm 0.07 \;, \label{f1}
\ee
\noindent which, using $\Gamma(\phi)=4.458$ MeV \cite{pdg} and
Eq.~(\ref{rate}), gives:
\be
{\cal B}(\phi \to f_0 \gamma)  =(2.7 \pm 1.1)~10^{-4} \; .
\label{brf0}
\ee
Both the results Eq.~(\ref{ftris}) and  Eq.~(\ref{f1})
have been derived without the inclusion of radiative
$\alpha_s$ corrections, an approximation which is usually believed more
accurate for the three point sum rule, where the ${\cal O}(\alpha_s)$
corrections are expected to cancel in the ratio of a three-point  and
a two-point function. 

Our result of  Eq.~(\ref{brf0}) is in reasonable agreement with  the outcome 
of refs. \cite{lucio,closebook}, where  the decay is supposed to
proceed through the chain $\phi \to K {\overline K} \gamma \to f_0
\gamma$,  and so depends on the coupling $ g_{f_0 K{\overline K}}$
\footnote{Such a coupling is taken from ref. \cite{truong}.}. Their 
results
are: ${\cal B}(\phi \to f_0 \gamma)=1.9 \times 10^{-4}$ \cite{lucio} and
${\cal B}(\phi \to f_0 \gamma)=1.35 \times 10^{-4}$
\cite{closebook}.
On the other hand, QCD spectral sum rules are exploited in  ref.
\cite{narison} to predict
${\cal B}(\phi \to f_0 \gamma)=1.3 \times 10^{-4}$.

A different strategy is proposed  in \cite{twophoton}, where
the experimental datum is assumed together with the structure
$f_0(980)=n{\overline n} \cos \theta +s {\overline s}\sin \theta$, where $n{\overline n} = (u{\overline u} + d{\overline d})/\sqrt{2}$ and
$\theta$ is a mixing angle. A theoretical prediction is derived describing
the particles ($\phi$, $f_0$) through wave functions depending on the
radii of the mesons. Such a prediction is then compared to the
experimental datum in order to constrain the mixing angle.

Although our result of Eq.~(\ref{brf0}) is
affected by a rather large uncertainty, it is in
agreement with the available  data \cite{akh,snd}. Since our
sum rule analysis is based on the hypothesis that the $f_0(980)$ couples to the
scalar ${\overline s}s$ current, this agreement
leads to the conclusion that an ${\overline s}s$
component is present in such a state. 
However, our branching ratio is an order of
magnitude larger than the naive quark model gives for a pure 
${\overline s}s$ state. Our result is consequently 
consistent with the view that 
the $f_0(980)$ is a meson with a basic  ${\overline q}q$ composition, which 
spends a sizeable part of its lifetime in a two meson state, 
such as $K{\overline K}$. This is in keeping with the 
analyses of \cite{tornqvist,scadron,boglione,torn2000} that attribute
such multi-hadron components  to dressing.
While the effect of
$K{\overline K}$ couplings have been studied phenomenologically
in Ref. \cite{twophoton} in a range of hadronic reactions,
 they have been dynamically calculated by Marco et al. \cite{marco}
explicitly for the radiative decay we study here and found
to give ${\cal B}(\phi\to f_0\gamma) = 2.4
\times 10^{-4}$,  in reassuringly good agreement with our sum-rule result,
Eq. (\ref{brf0}).


\vspace{1cm}
We acknowledge the EU-TMR Programme, Contract No. CT98-0169,
EuroDA$\Phi$NE for support. F.D.F. warmly thanks the Institute
for Particle Physics Phenomenology, University of Durham, where
this work was started.


\begin{thebibliography}{99}

\bibitem{rassegne}
For reviews see: 
L. Montanet, 
Rep. Prog. Phys. {\bf 46} (1983) 337; 
F.E. Close, 
Rep. Prog. Phys. {\bf 51}(1988) 833; 
N.N. Achasov, 
Nucl.Phys. Proc. Suppl. {\bf B21}  (1991) 189; 
T. Barnes, 
hep-ph/0001326.

\bibitem{isgur}
J. Weinstein and N. Isgur, 
Phys. Rev. Lett. {\bf 48} (1982) 659; 
Phys. Rev. {\bf D27} (1983) 588; 
Phys. Rev. {\bf D41} (1990) 2236.

\bibitem{au}
K.L. Au, D. Morgan and M.R. Pennington, 
Phys. Rev. {\bf D35} (1987) 1633.

\bibitem{janssen}
G. Janssen {\it et al.}, 
Phys. Rev. {\bf D52} (1995) 2690.

\bibitem{oller}
J.A. Oller and E. Oset, 
Phys. Rev. {\bf D60} (1999) 074023; 
Nucl. Phys. {\bf A620} (1997) 438, Nucl. Phys. {\bf A652} (1999) 407 (E).

\bibitem{tornqvist} 
N.A. Tornqvist, 
Phys. Rev. Lett. {\bf 49} (1982) 624;
Z. Phys. {\bf C68} (1995) 647.      

\bibitem{roos}
N.A. Tornqvist and M.Roos, 
Phys. Rev. Lett. {\bf 76} (1996) 1575.

\bibitem{scadron}
E. van Beveren {\it et al.}, 
Z. Phys. {\bf C30} (1986) 615; 
M.D. Scadron, 
Phys. Rev. {\bf D26} (1982) 239; 
E. van Beveren, G. Rupp and M.D. Scadron, 
Phys. Lett. {\bf B495} (2000) 300.

\bibitem{jaffe}
R.L. Jaffe, 
Phys. Rev. {\bf D15} (1977) 267, 281; 
{\bf D17} (1978) 1444;
R.L. Jaffe and K. Johnson, 
Phys. Lett. {\bf B60} (1976) 201.


\bibitem{ivan}
N.N. Achasov and V.N. Ivanchenko, 
Nucl. Phys. {\bf B315} (1989) 465; 
N.N. Achasov and V.V. Gubin, 
Phys. Rev. {\bf D56} (1997) 4084.


\bibitem{closebook}
N. Brown and F.E. Close, 
in ref. \cite{book}, pp. 447-464;
F.E. Close, N. Isgur and S. Kumano,
Nucl. Phys. {\bf B389} (1993) 513.

\bibitem{kaminski}
R. Kaminski, L. Lesniak and J.P. Maillet, 
Phys. Rev. {\bf D50} (1994) 3145.

\bibitem{shev} 
N.N. Achasov, V.V. Gubin and V.I. Shevchenko,
Phys. Rev. {D56} (1997) 203.

\bibitem{morgan}
D. Morgan and M.R. Pennington, 
Phys. Lett {\bf B258} (1991) 444;
ibidem {\bf B269} (1991) 477 (E). 

\bibitem{pdg}
{\it Review of Particle Physics}, D.E. Groom et al.,
Eur. Phys. J. {\bf C15} (2000) 1.

\bibitem{twophoton}
A.V. Anisovich, V.V. Anisovich and V.A. Nikonov, 
hep-ph/0011191.

\bibitem{boglione}
M. Boglione and M. R. Pennington, 
Phys. Rev. Lett. {\bf 79} (1997) 1998.  

\bibitem{akh}
R.R. Akhmetshin et al., 
Phys. Lett. {\bf B462} (1999) 380.

\bibitem{snd}
M.N. Achasov et al., 
Phys. Lett. {\bf B440} (1998) 442.

\bibitem{book}
For a review, see: {\it The DA$\Phi$NE Physics HandBook}, L. Maiani,
G. Pancheri and N. Paver eds, INFN Frascati, 1995.

\bibitem{kloe} 
KLOE Collab., M. Adinolfi {\it et al.}, 
hep-ex/0006036.     

\bibitem{shifman}
M.A. Shifman, A.I. Vainshtein and V.I. Zakharov,
Nucl. Phys. {\bf B147} (1979) 385, 448. 
For a review on the QCD sum-rule
method see  {\it Vacuum Structure and QCD Sum Rules},
M.A. Shifman ed., North-Holland, Amsterdam, 1992.

\bibitem{noi}
F. De Fazio and M.R. Pennington, 
JHEP {\bf 0007} (2000) 051.

\bibitem{liu}
R. Delbourgo, D.-S. Liu and M.D. Scadron, Phys. Lett. {\bf B446} (1999)
332.

\bibitem{gidal}
G. Gidal et al., MARK II Collab., Phys. Lett. {\bf B107} (1981) 153;
U. Malik, MARK III Collab., Proc. XXI Rencontre de Moriond, Vol. 2, ed J.
Tran  Thanh Vanh (Editions Fronti\`eres, 1986), p. 431;
 W. Lockman, Mark III Collab., Proc. 3rd Int. Conf. on Hadron Spectroscopy
(Ajaccio, France), eds F Binon et al. (Editions Fronti\`eres, 1989)
  p.109;  A. Falvard et al., DM2 Collab., Phys Rev {\bf D38} (1988) 2706.

\bibitem{anjos}
J.C. Anjos et al., E691 Collab., Phys. Rev. Lett. {\bf 62} (1989) 125 and
more recent  E.M. Aitala et al. E791 Collab., Phys. Rev. Lett. {\bf 86}
(2001) 765.

\bibitem{morgan1}
D. Morgan and M.R. Pennington, Phys. Rev. {\bf D48} (1993) 1185. 

\bibitem{noistrano}
P. Colangelo, F. De Fazio, G. Nardulli and N.Paver, 
Phys. Lett. {\bf B408} (1997) 340. 
For other determinations see the review:
P. Colangelo and A. Khodjamirian, 
hep-ph/0010175.


\bibitem{lucio}
J. Lucio and J. Pestieau, 
Phys. Rev. {\bf D42} (1990) 3253.

\bibitem{truong}
S. Nussinov and T.N. Truong, 
Phys. Rev. Lett. {\bf 63} (1989) 1349; 2002 (E).

\bibitem{narison}
S. Narison, 
Nucl. Phys. Proc. Suppl. {\bf 96} (2001) 244.

\bibitem{torn2000}
N.A. Tornqvist,
hep-ph/0008136.

\bibitem{marco}
 E. Marco et al, Phys. Lett. {\bf B470} (1999) 20. 

\end{thebibliography}
\end{document}